ORIGINAL PAPER

# Human mobility and COVID-19 initial dynamics

**Stefano Maria Iacus** · **Carlos Santamaria** · **Francesco Sermi** · **Spyros Spyratos** · **Dario Tarchi** · **Michele Vespe**

**Abstract** Countries in Europe took different mobility containment measures to curb the spread of COVID-19. The European Commission asked mobile network operators to share on a voluntarily basis anonymised and aggregate mobile data to improve the quality of modelling and forecasting for the pandemic at EU level. In fact, mobility data at EU scale can help understand the dynamics of the pandemic and possibly limit the impact of future waves. Still, since a reliable and consistent method to measure the evolution of contagion at international level is missing, a systematic analysis of the relationship between human mobility and virus spread has never been conducted. A notable exceptions are France and Italy, for which data on excess deaths, an indirect indicator which is generally considered to be less affected by national and regional assumptions, are available at department and municipality level, respectively. Using this information together with anonymised and aggregated mobile data, this study shows that mobility alone can explain up to 92% of the initial spread in these two EU countries, while it has a slow decay effect after lockdown measures, meaning that mobility restrictions seem to have effectively contribute to save lives. It also emerges that internal mobility is more important than mobility across provinces and that the typical lagged positive effect of reduced human mobility on reducing excess deaths is around 14–20 days. An analogous analysis relative to Spain, for which an IgG SARS-Cov-2 antibody screening study at province level is used instead of excess deaths statistics, confirms the findings. The same approach adopted in this study can be easily extended to other European countries, as soon as reliable data on the spreading of the virus at a suitable level of granularity will be available. Looking at past data, relative to the initial phase of the outbreak in EU Member States, this study shows in which extent the spreading of the virus and human mobility are connected. The findings will support policymakers in formulating the best data-driven approaches for coming out of confinement and mostly in building future scenarios in case of new outbreaks.

## 1 Introduction

By means of a letter sent on 8 April 2020, the European Commission asked European mobile network operators (MNOs) to share fully anonymised aggregate mobility data to deliver insights to help fight COVID-19. The aim of the initiative is further specified by the recommendation to support exit strategies policies through mobile data and apps [5] with data-driven models that indicate the potential effects of the relaxation of the physical distancing measures [6].

S. M. Iacus (✉)· C. Santamaria · F. Sermi · S. Spyratos · D. Tarchi · M. Vespe
Joint Research Centre, Via Enrico Fermi 2749, 21027 Ispra, VA, Italy
e-mail: stefano.iacus@ec.europa.eu

Indeed, MNOs can offer a wealth of data on mobility and social interactions. While the value of mobile positioning *personal data* to describe human mobility has been explored [2] and its potential in epidemiology studies demonstrated [16,17,25–27], the aim of this research is to show that also anonymised and aggregated MNOs data, in compliance with the 'Guidelines on the use of location data and contact tracing tools in the context of the COVID-19 outbreak' by the European Data Protection Board EDPB [3], could serve to explain the dynamics of the virus outbreak without the need of disclosing any personal data, nor using tracing applications.

The analysis in this study exploits data from four MNOs relatively to three EU Member States (France, Italy and Spain), together with different types of official statistics on the spread of COVID-19 across these countries; the objective is to understand whether or not human mobility matters in the expansion phase of the pandemic.

The paper is organised as follows: Sect. 2 briefly explains the nature of the mobile data used in this work, while Sect. 3 investigates the case of France, taking into account the internal and outbound movements from the Haut-Rhin department (one of the initial hotbeds) to the rest of the country, in order to establish the role of human mobility in explaining the dynamics of the early spread of the virus. The result show that significantly higher excess mortality in other departments can be explained by their connectivity to Haut-Rhin ($R^2$ up to 0.92). This relationship is valid until mobility is severely reduced due to the national lockdown in force since the 16 March 2020. It also finds out that the curve of cumulative excess deaths in Haut-Rhin correlates almost perfectly (from $R^2 = 0.69$ to $R^2 = 0.98$ after shifting) with the curve of mobility reduction for the same region, showing a lag of 14 days between the two curves.

Section 4 focuses on the virus spread in Italy, showing rather similar results to the French case ($R^2$ up to 0.91 and lagged effect of 14–20 days, depending on the provinces) and confirming that human mobility has an high impact on the virus spread. Since the excess mortality data for Italy is available at municipality level, we have tested the same method also at this higher spatial granularity (the analysis for France is at Departments level), finding an analogous relationship between mobility and excess deaths in the initial phase of the outbreak; this confirms that the results hold at different geographical scales.

Section 5, focuses on Spain, for which instead of the excess death statistics (that are not available at a suitable spatial granularity) we consider the *IgG SARS-Cov-2 antibody screening study* put in place at province level between the end of April and mid-May 2020. In this case, only correlation can be tested, and the results show that the number of people found to be positive to IgG test is highly correlated ($\rho$ up to 75%) with the human mobility derived by the mobile data from the MNOs. It is worth underlining that since, as noticed in Salje et al. [21], population immunity appears insufficient to avoid a second wave without lockdown or other containment measures, this work may provide further insights on the effectiveness of containment measures.

Section 6 goes through the main caveats and limitation of the research, pointing out the hypothesis adopted by the proposed analysis.

Finally, Sect. 7 draws some considerations about the research and highlights the main findings.

## 2 Mobile positioning data and mobility indicators

The agreement between the European Commission and the mobile network operators (MNOs) defines the basic characteristics of fully anonymised and aggregate data to be shared with the Commission's Joint Research Centre (JRC[1]). The JRC processes the heterogeneous sets of data from the MNOs and creates a set of mobility indicators and maps at a level suitable to study mobility comparatively across countries; this level is referred to as 'common denominator'.

The following subsections briefly describe the original mobile positioning data from the MNOs and the mobility indicator derived by JRC, respectively.

### 2.1 Origin–destination matrix

Data from MNOs are provided to JRC in the form of origin–destination matrices (ODMs) [7,20]. Each cell $[i - j]$ of the ODM shows the overall number of *'movements'* (also referred to as 'trips' or 'visits') that have been recorded from the origin geographical reference area $i$ to the destination geographical reference area $j$ over the reference period. In general, an ODM is structured as a table showing:

- reference period (date and, eventually, time);
- area of origin;
- area of destination;
- count of movements.

[1] The Joint Research Centre is the European Commission's science and knowledge service. The JRC employs scientists to carry out research in order to provide independent scientific advice and support to EU policy.

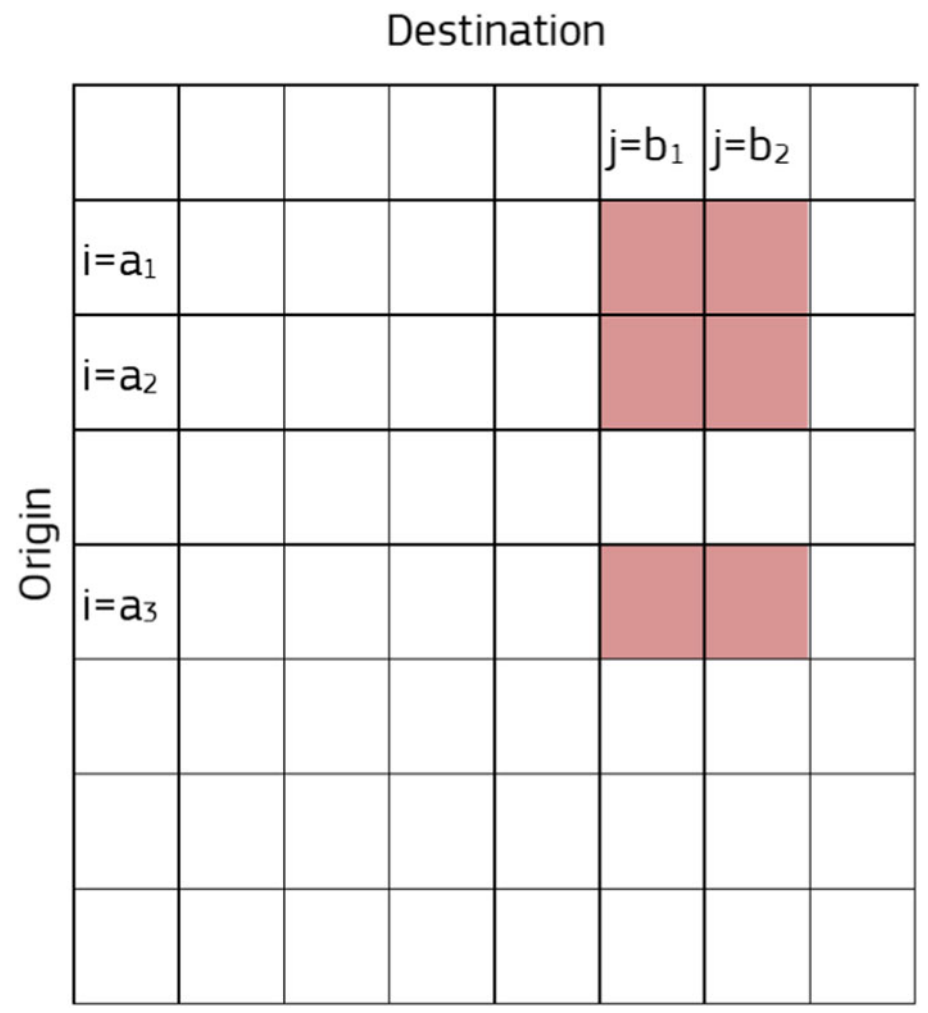

**Fig. 1** Connectivity matrix construction starting from an ODM at higher level of granularity

Despite the fact that the ODMs provided by different MNOs have similar structure, they are often very heterogeneous. Their differences can be due to the methodology applied to count the movements, to the spatial granularity or to the time coverage. Nevertheless, each ODM is consistent over time and relative changes are possible to be estimated. This allows defining common indicators (such as 'mobility indicators' [22], 'connectivity matrices' and 'mobility functional areas' [13] that can be used, with all their caveats, by JRC in the framework of this joint initiative.

Although the ODM contains only anonymised and aggregate data, in compliance with the EDPB guidelines [3], upon the reception of each ODM, the JRC carries out a *'Reasonability Test'*. Both the reasonability test and the processing of the ODM to derive mobility indicators take place within the JRC's *Secure Platform for Epidemiological Analysis and Research* (SPEAR).

This study presents an exploratory analysis for specific outbreaks in France, Italy and Spain, showing and quantifying the relative importance of mobility in the early stages of the virus outbreak in the country. The choice of these three countries is attributed to the availability of ancillary data at high level of granularity in space and time, which are needed for correlation analysis with mobility insights. The mobility within the country is derived using the 'connectivity matrices' that are introduced in the following section.

### 2.2 Connectivity matrix

Besides the formulation of 'mobility indicators' [22] and 'mobility functional areas' [13], an additional example of 'common denominator' to guarantee comparable indicators across countries is a measure of *connectivity* at less granular level, using the common statistical classification of territorial units (NUTS[2]). The NUTS3 level was selected in this framework, where the average population size of administrative units in Member States lie between 150.000 and 800.000.

The connectivity matrix at time $t$ is represented by the origin–destination matrix $ODM_{i,j}(t)$ as extracted from normalised and aggregated data received from MNOs. With reference to Fig. 1, considering the NUTS3 area $A$ (rows and columns $a1$, $a2$ and $a3$) and the destination area $B$ (rows and columns $b1$, $b2$), then the relevant connectivity is calculated as follows:

$$\mathrm{ODM}_{A,B}(t) = \sum_{i \in \{a_1,a_2,a_3\}} \sum_{j \in \{b_1,b_2\}} \mathrm{ODM}_{i,j}(t) \qquad (1)$$

The resulting connectivity matrix is obtained by averaging over 1 week of data. The present study only makes use of outward and internal movements along one row of the matrix and later focuses only on movements internal to departments, i.e. making use only of the diagonal of the connectivity matrix.

## 3 Human mobility explains the early spread of COIVD-19 in France

One of the initial clusters of the COVID-19 outbreak in France is believed to have originated in Haut-Rhin. Since statistics on confirmed cases are by their nature

[2] Regulation (EC) 1059/2003 of the European Parliament and of the Council of 26 May 2003 on the establishment of a common classification of territorial units for statistics (NUTS)

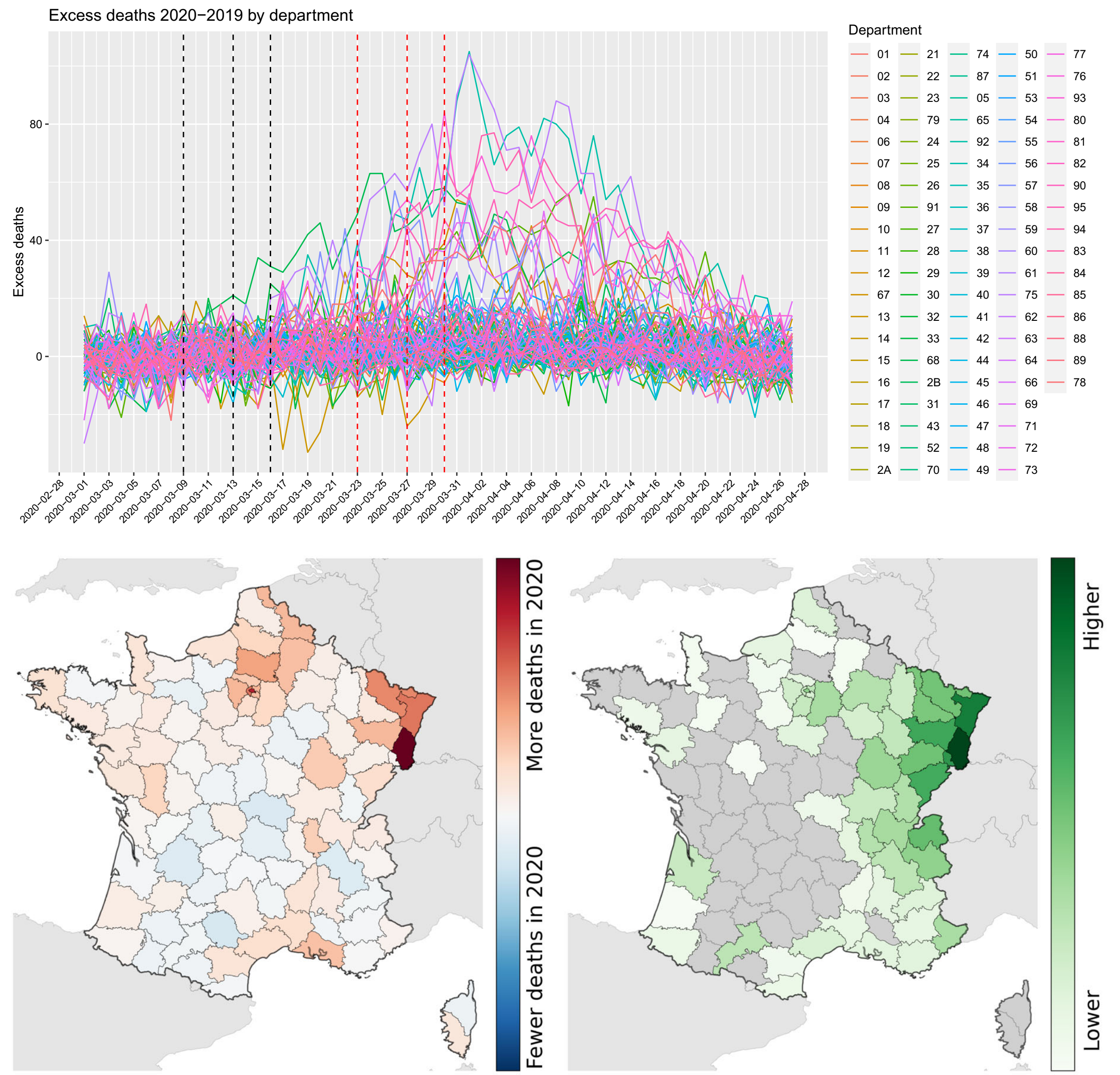

**Fig. 2** Top panel: Evolution of the excess deaths by department, 1 March–27 April 2020. Bottom left: Cumulative number of excess deaths 2020–2019, period 1–25 March. Bottom right: Connectivity levels from the department of Haut-Rhin (darkest area in the map) during the week of 23–29 February 2020. Grey colour indicates very low connectivity

heavily affected by the type and volume of the testing procedures [1], which generally varies both across departments and in time as the contagion evolves, we instead have used the total number of (officially confirmed) deaths by departments (see Fig. 2 left) over the period 1 March 2020–27 April 2020 [14]. France is a special case study thanks to the availability of high-resolution deaths counts and our results seem to confirm and complement the analysis in [21]. In our analysis, we consider the daily cumulative number of excess deaths with respect to the same period of 2019, assuming that most of these excess cases are due to COVID-19 [8,23,28]. A visual inspection of the data shows that the cumulative sum of excess deaths by department (Fig. 2 middle) for the period 1–25 March 2020 is correlated with the connectivity (Fig. 2 right) from Haut-Rhin dur-ing the week of 23–29 February 2020 (which is supposed to be the start of the pandemic in France). To study the impact of mobility on the spread of COVID-19 in France, we fit a simple statistical model that aims at explaining the excess deaths of Haut-Rhin in terms of human mobility as well as the distance between the department of Haut-Rhin and all other departments. We aim to measure the relative importance of connectivity (or human mobility) compared to the geographical distance for the spread of the virus.

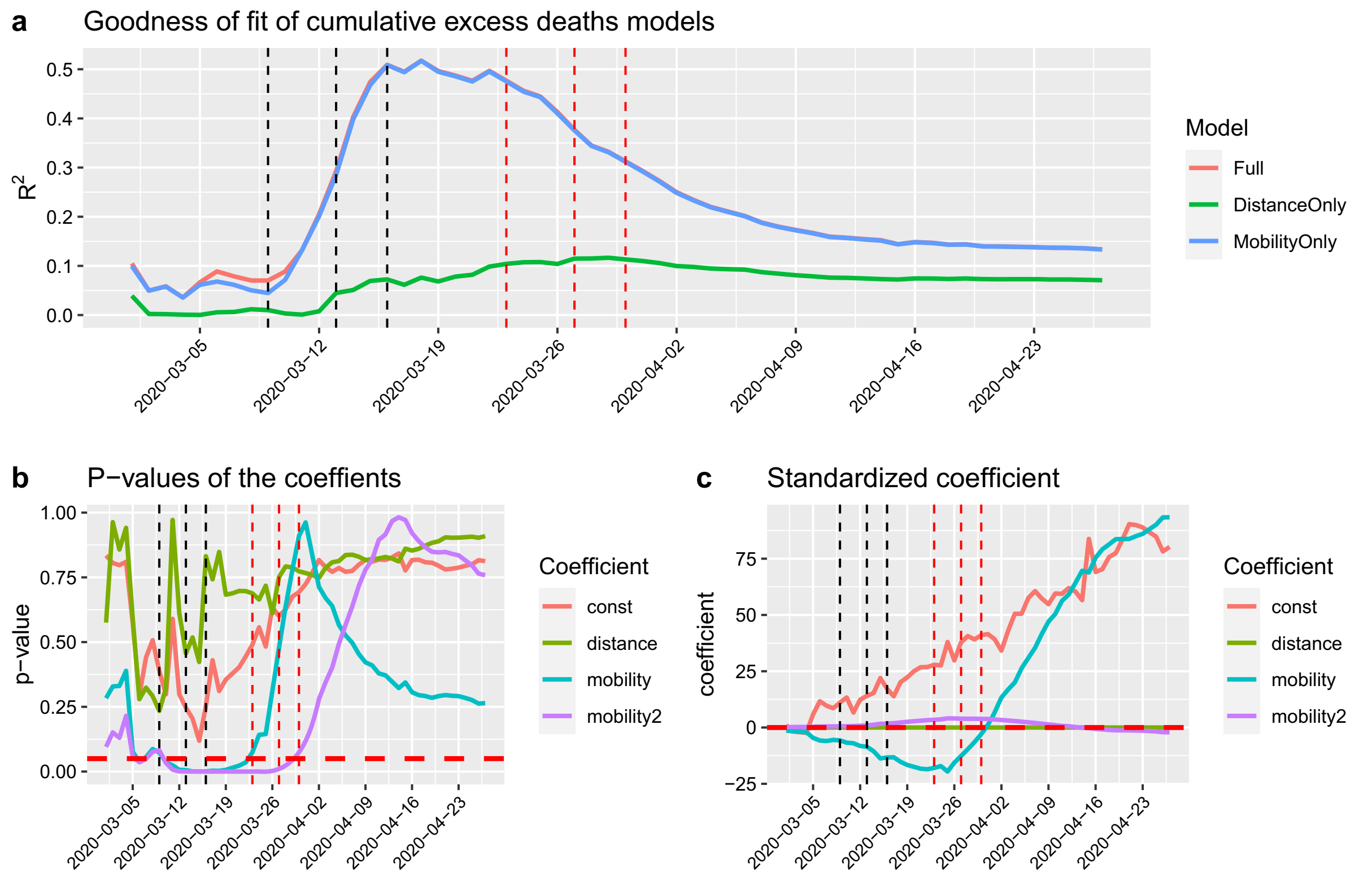

**Fig. 3** **a** The $R^2$ is explained mostly by the mobility indicator and reaches its maximum on 16 March 2020. When the lockdown measures are in force, then the spread is explained by the distance from the Haut-Rhin department, where the spreading of the virus has believed to have started, but also other factors captured by the constant effect. **b** The $p$ values of the coefficients are presented. The horizontal line represents the 0.05 significance level and **c** the standardised values of the same coefficients to appreciate the relative impact of each on explaining the outcome variable. (Color figure online)

Although mobility can only explain part of the spread of COVID-19, in order to show the relative importance of the mobility component, we run the simple quadratic regression model of equation:

$$y_i^d = const + \alpha_1 \log(\text{mobility}_i) + \alpha_2 \left[\log(\text{mobility}_i)\right]^2 + \alpha_3 \text{distance}_i \quad (2)$$

where $y_i^d$ is the number of cumulative death excess for department $i$ from 1 March 2020 till date $d$ ($d$ = 2020-03-01, …, 2020-04-27). The variable distance$_i$ is the geographical distance between the department Haut-Rhin and the department $i$ and mobility$_i$ is the normalised Mobility Index from Haut-Rhin to the department $i$ for the week of 23–29 February 2020. As we consider log-scaled mobility, we drop the departments for which the outbound mobility from Haut-Rhin is zero, i.e. we only fit this and the other models for which mobility$_i > 0$. The idea behind this data-driven approach is to let the data tell us under which conditions and when the mobility data do matter in studying the contagion and when they do not. The choice of the very simple structure of model (2) is to capture clearly the impact of mobility, and the scope of the model is not to forecast cases or fatalities, but to inform about the dynamics of the COVID-19 due to human mobility. The quadratic form of the model is suggested by the data itself as shown in Fig. 4. We also fit two reduced models with *mobility* only (3) and *distance* only (4):

$$y_i^d = \text{const} + \beta_1 \log(\text{mobility}_i) + \beta_2 \left[\log(\text{mobility}_i)\right]^2 \quad (3)$$

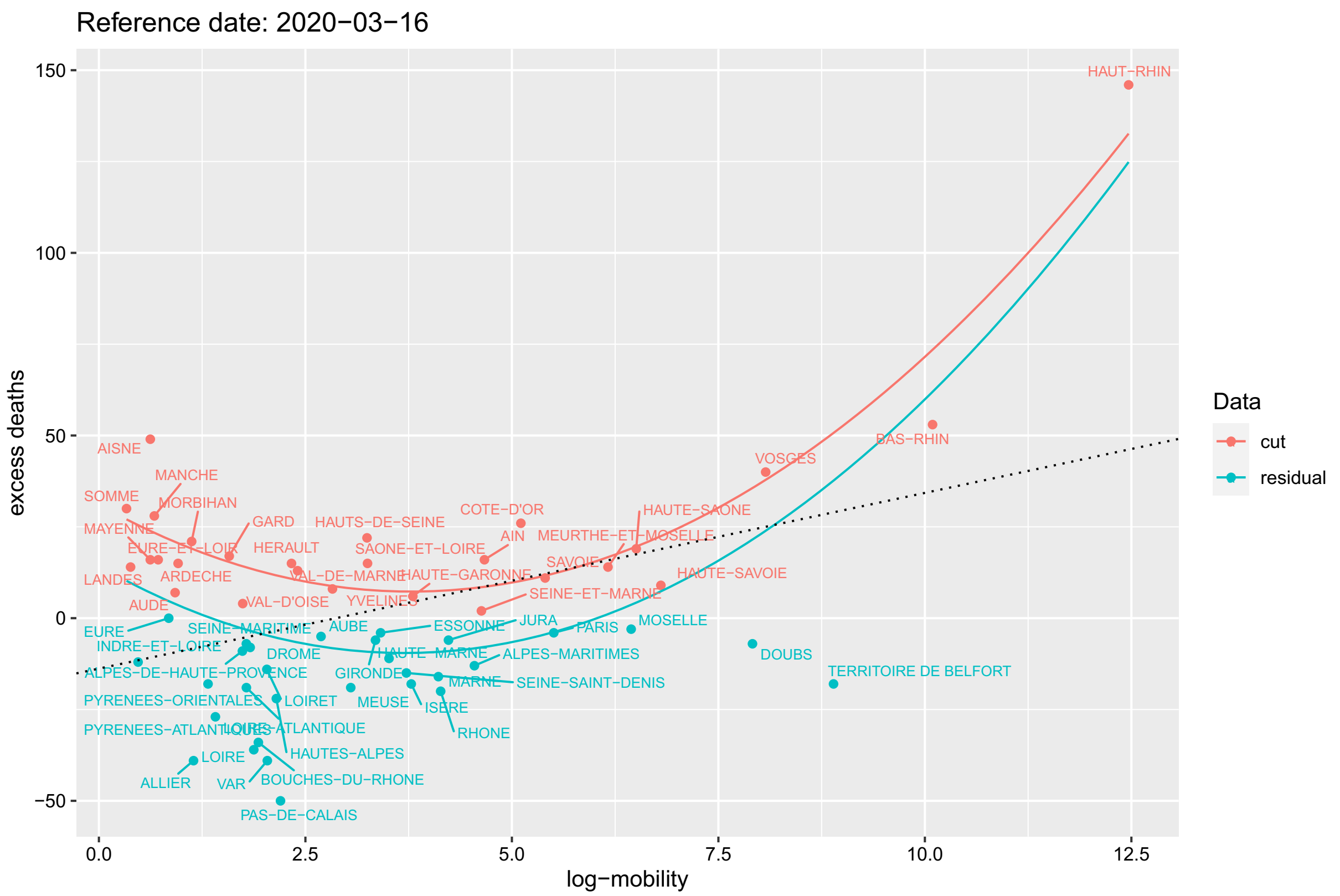

**Fig. 4** Data for log-mobility and cumulative excess deaths on 16 March 2020. Red represents the selected subset of positive excess deaths and mobility. The other points are the residual data. Red and light blue together represent the whole data. The two curves correspond to the fitted model (2) for both data sets. The straight line represent simple linear regression of the two variables in the plot. (Color figure online)

and

$$y_i^d = \text{const} + \gamma_1 \text{distance}_i . \tag{4}$$

For each model, we evaluate the $R^2$ index as well as the significance of each coefficient. To further stress the relative importance of the variables in the full model (2), we consider the evolution of the standardised coefficients in time. Figure 3 shows the goodness of fit of the three competing models in terms of $R^2$ and the $p$ value of the estimated coefficients. The vertical lines set on 9 and 13 March represent two large gatherings in the Haut-Rhin region and that on 16 March the first lockdown measure (schools closure). The next three lines are translated by 14 days, which corresponds to the median number of days from the occurrence of the first symptom to death [18]. Figure 3 shows that the model (2) dominates the other two but also that, up to the maximum value of the $R^2 = 0.52$ (around 18 March), the model (3) is almost equally good ($R^2 = 0.51$). Then, on the long run, the distance becomes slightly more important, but clearly the rest of the variability is explained by the mix of the three. This figure also supports an argument that the department of Haut-Rhin could have been the initial source of the outbreak of COVID-19 in France. It also demonstrates the positive impact of the lockdown measures on the reduction of excess deaths. Figure 3 also shows the standardised coefficients of the full model (2) as a function of time. Standardised coefficients are all on the same scale, and thus, their values can be compared in terms of their impact on the outcome variable. Looking at the plot, then it becomes clear that the mobility coefficient is the largest effect up to 27 March; then, the excess death number is explained by the constant in the model, i.e. by other factors not included in the simple regression analysis.

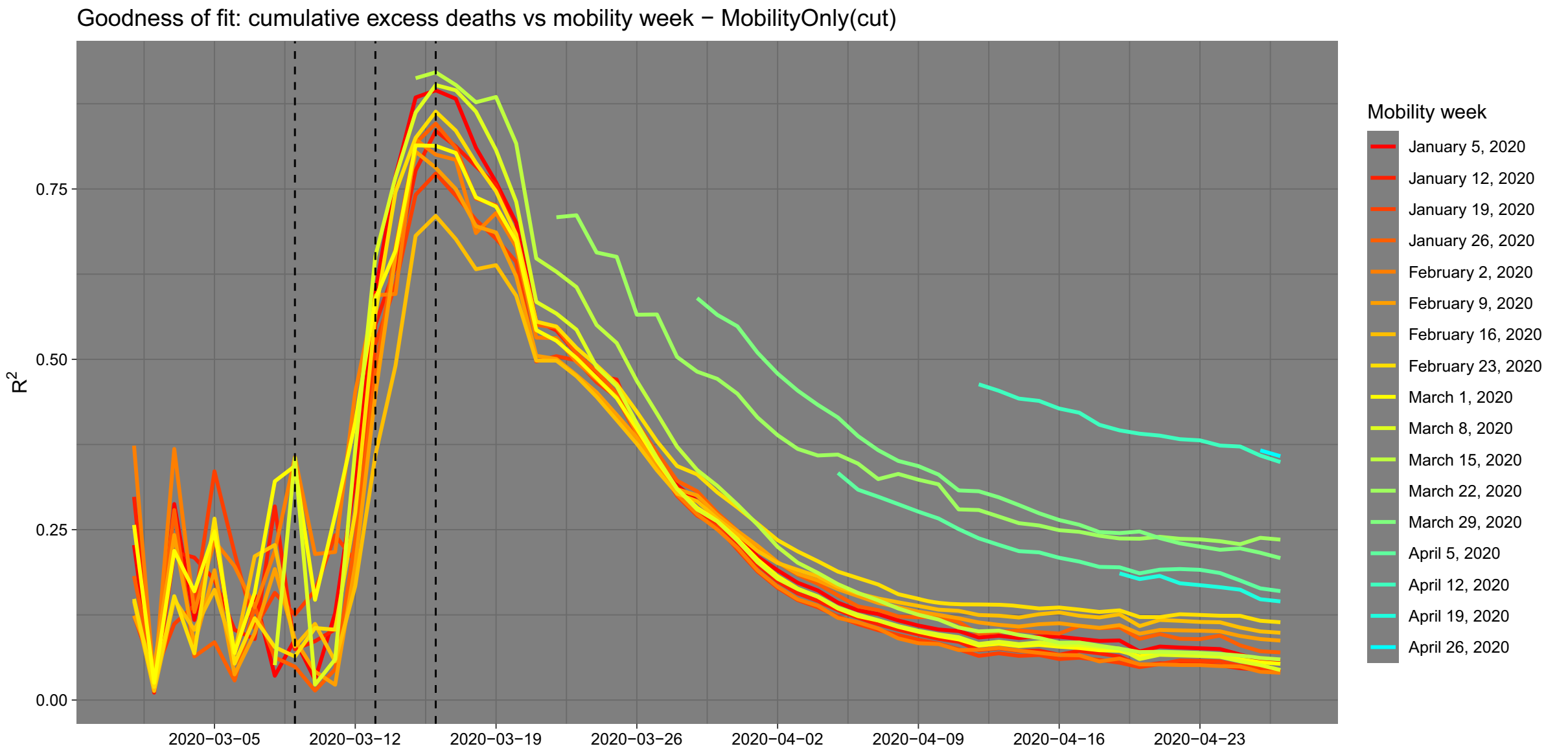

**Fig. 5** $R^2$ evolution for each mobility week using the model based on mobility alone on the selected data set of positive excess deaths and mobility index. The peak of correlation is on 16 March 2020 and the mobility week is the one between 15 and 21 March 2020. A similar pattern is shown for all weeks, explaining that mobility has no effect before the COVID-19 outbreaks and has a fast decay after the lockdown measures are in place (left and right tails of the figure). On the contrary, during the initial emergency phase mobility is a key element of the spread of the virus and therefore on fatalities

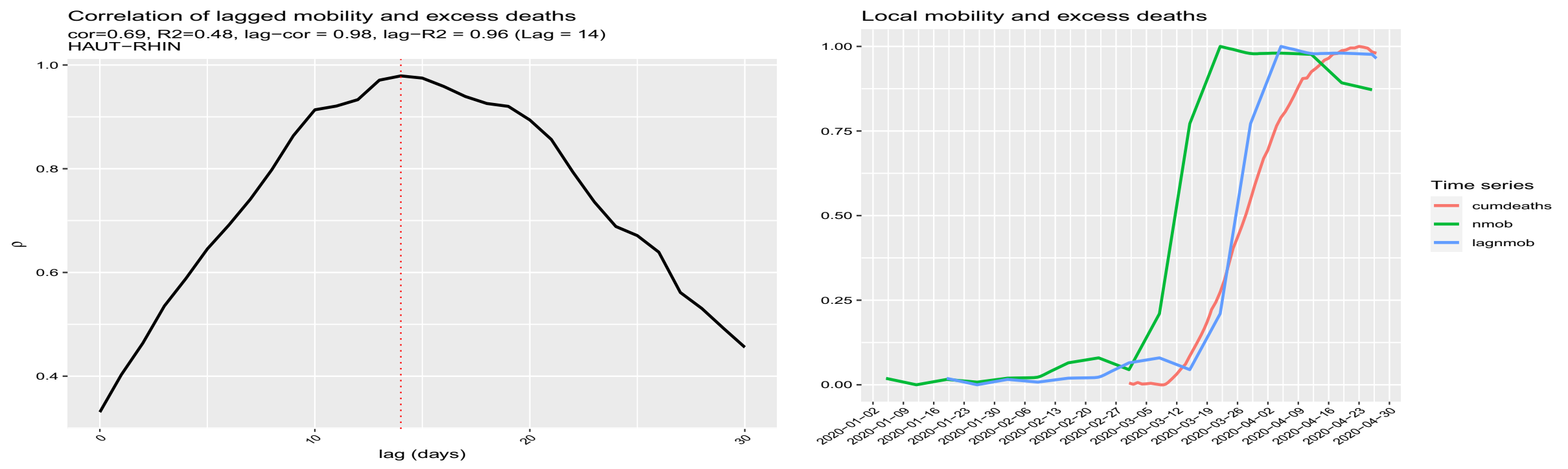

**Fig. 6** Left panel: Correlation between `cumdeaths` and `nmob` at different lags. Right panel: `cumdeaths`, `nmob` with the additional lagged normalised mobility curve `lagnmob`. The estimated optimal lag is 14 days

The explanatory impact of mobility can be improved further if we select those cases for which the mobility index and the cumulative excess deaths are *both* positive. This selection is natural both because what is interesting is the positive excess of deaths and because, due to the process of non-identification and anonymisation of the data, some entries of the original OD matrix are cut to zero by construction. Figure 4 shows the scatterplot of the two dimensions for all the departments and the further selection of data (`cut`). The reference data chosen is 16 March as it is the date for which the $R^2 = 0.92$ of the full model, under the selected sample, is maximal. For the selected data, the significance of the coefficients for the constant and the mobility variable are even improved, while for the variable distance is worsened and the standardised coefficients show a similar path to those of Fig. 3. Figure 5 also shows that this evidence does not change substantially if we choose a different mobility week. Indeed, the figure shows that the $R^2$ evolution for each mobility week using the model (3) on the selected data set (`cut`) of positive excess deaths and mobility index present a common pattern, i.e. mobility has no effect before the COVID-19 outbreaks and has a fast decay after the lockdown measures are in place (left and right tails of Fig. 5). On the contrary, during the initial emergency phase, mobility is a key element of the spread of the virus and therefore on its outcome (the fatalities). Although we are not trying to forecast the number of deaths with mobility due to the extreme simplicity of the models considered, we notice that the $R^2$ path shows an observed time lag between mobility and excess of deaths which we impute to COVID-19: according to Li et al. [19] the mean incubation period was 5.2 days (95% confidence interval 4.1–7.0), while according to Wang et al. [24] the median number of days from the occurrence of the first symptom to death is 14 (range 6–41) days. See also Lauer et al. [18]. We now try to estimate this lag looking at internal mobility.

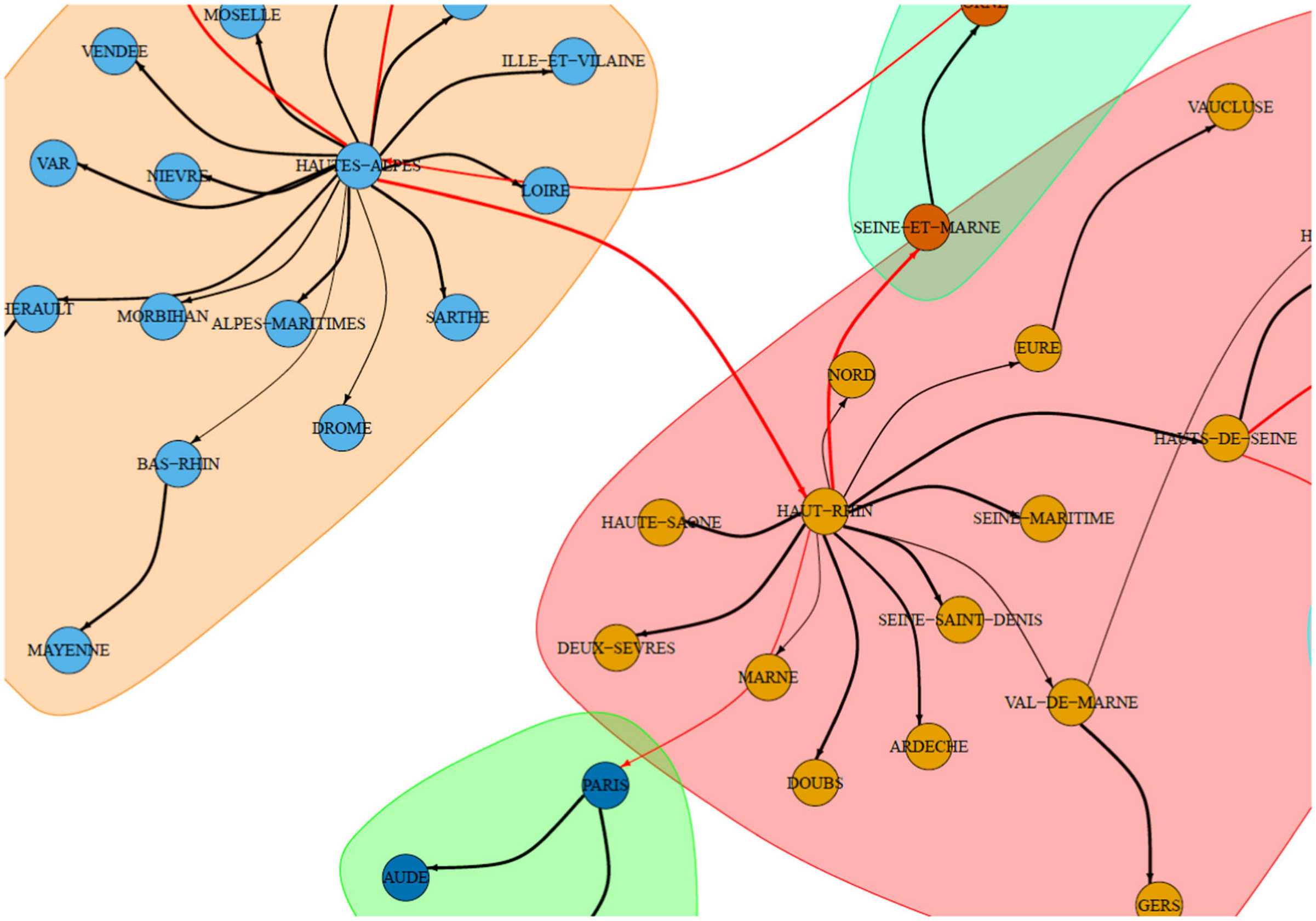

**Fig. 7** A zoomed section of the graph of lead–lag relationships between internal mobility (origin) and cumulative excess deaths (destination). Full network is shown in Fig. 8. The graph is zoomed on Haut-Rhin. This cluster shows that the internal mobility of Haut-Rhin correlates maximally with the cumulative excess deaths of several other departments at different lags. The width of the edges is inversely proportional to the estimated lag: the smaller the lag, the thicker the edge. Red edges connect two different clusters. No causality effect should be read from this graph. (Color figure online)

### 3.1 Internal mobility matters the most

Most of the correlation captured by the $R^2$ in the previous analysis is due to the internal mobility within the department and by the connectivity between the Haut-Rhin department and few others (Bas-Rhin, Vosges, Aisne, but not, for example, Moselle, see also Fig. 4). In fact, the distance alone cannot explain the excess deaths in the chosen simplified (2) model. Starting from this evidence, which is common to most of the departments considered in the analysis, we now

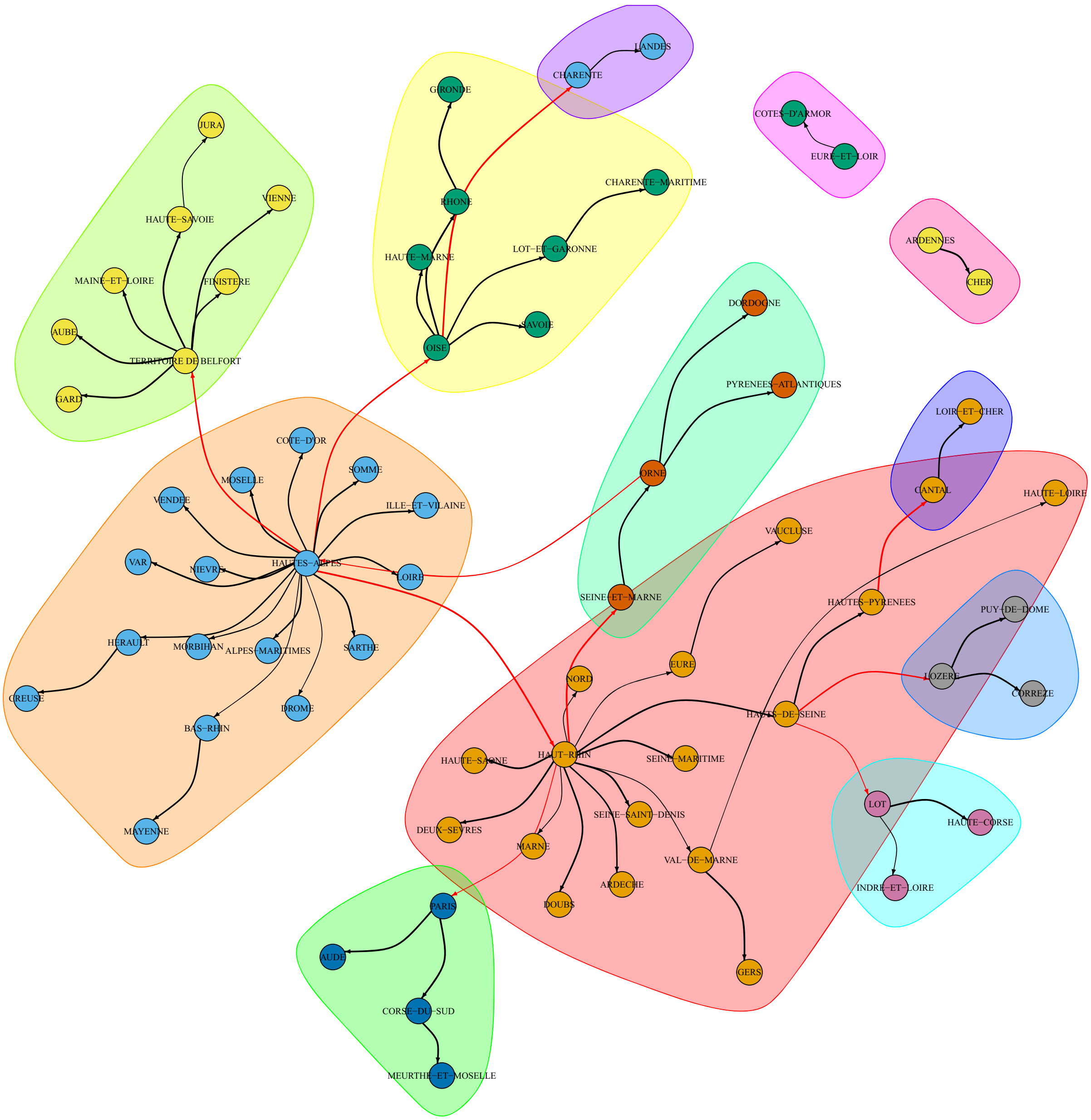

**Fig. 8** Full graph of lead–lag relationships and cluster of patterns. The width of the edges is inversely proportional to the estimated lag. Red edges connect two different clusters. No causality effect should be read from this graph. (Color figure online)

study the impact, if any, of internal mobility alone on the excess deaths of the same department through a time series approach. A similar study using domestic air traffic data at country level can be found in Zhao et al. [29]. In this work we consider the time series of excess deaths and the one of the reductions of mobility (100% = total lockdown, 0% no restriction to mobility) for each department through time. We normalise the reduction of mobility index (namely `nmob`), to the local maximum within the department to have a [0,1] measure

$$\text{nmob}_{i,j} = 1 - \frac{\text{mobility}_{i,j} - \min_j(\text{mobility}_{i,j})}{\max_j(\text{mobility}_{i,j}) - \min_j(\text{mobility}_{i,j})},$$

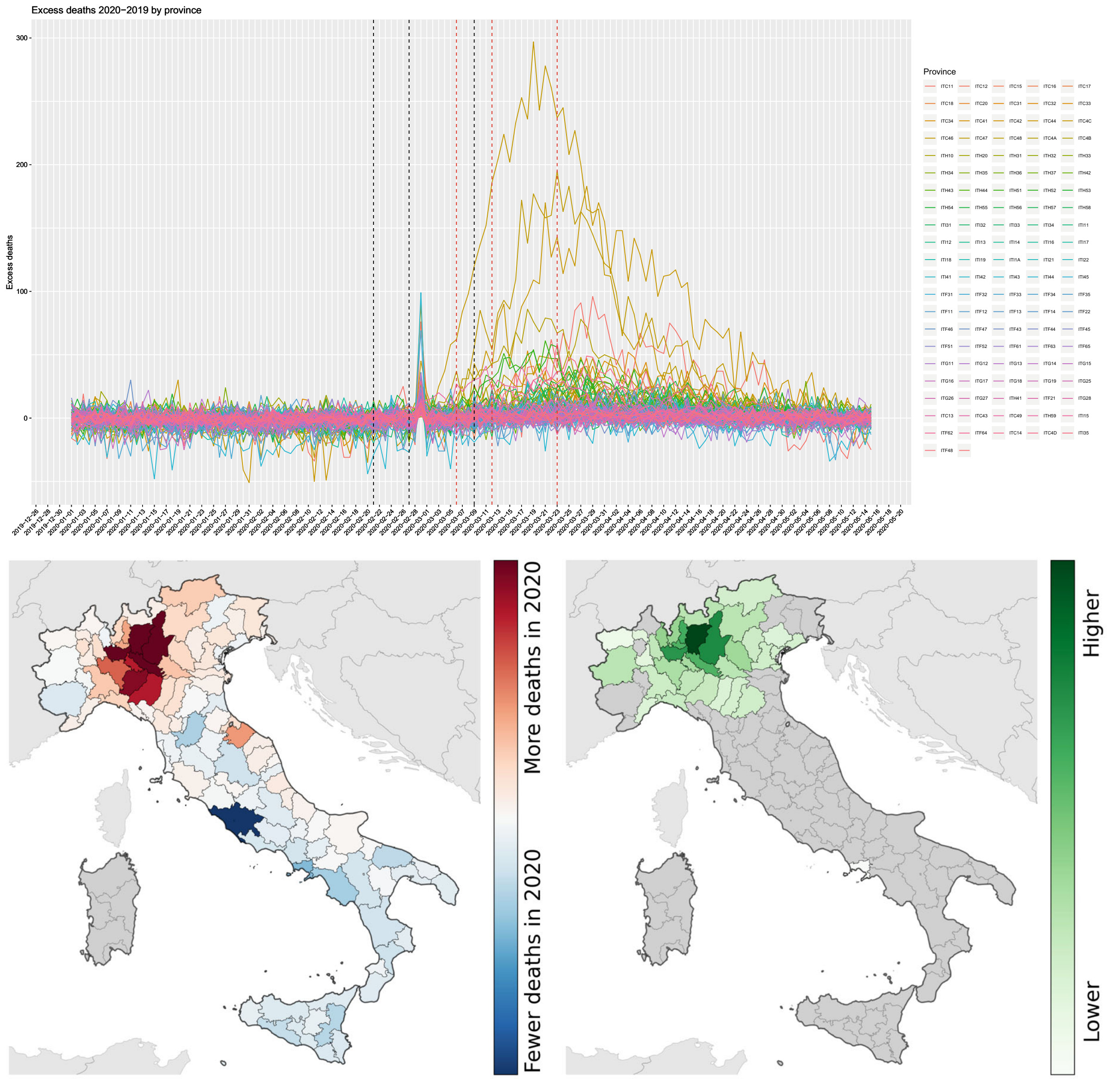

**Fig. 9** Top panel: Evolution of the excess deaths by province, 1 January–15 May 2020. Bottom left: Cumulative number of excess deaths 2020–2019, period 1 January–25 March. Bottom right: Connectivity levels from the province of Bergamo (darkest area in the map) during the week of 2 March 2020. Grey colour indicates very low connectivity

for each origin department $i$ and all outbound departments $j$. We also normalise the cumulative sums of excess deaths in the same way

$$\text{cumdeaths}_i = \frac{\text{cumdeaths}_i - \min_j(\text{cumdeaths}_i)}{\max_j(\text{cumdeaths}_i) - \min_j(\text{cumdeaths}_i)},$$

for each department $i$.

Figure 6 shows a plot of the two indicators for the Haut-Rhin department. The graphs suggest an evident lag could exist between the two curves. We can estimate the statistical lag in the following way. Let $\theta \in (-\delta, \delta)$ be the time lag between the two nonlinear time series $X$ and $Y$. Roughly speaking, the idea is to construct a contrast function $U_n(\theta) = \text{Cov}(X_t, Y_{t+\theta})$ which evaluates the Hayashi–Yoshida covariance estimator [9,10]

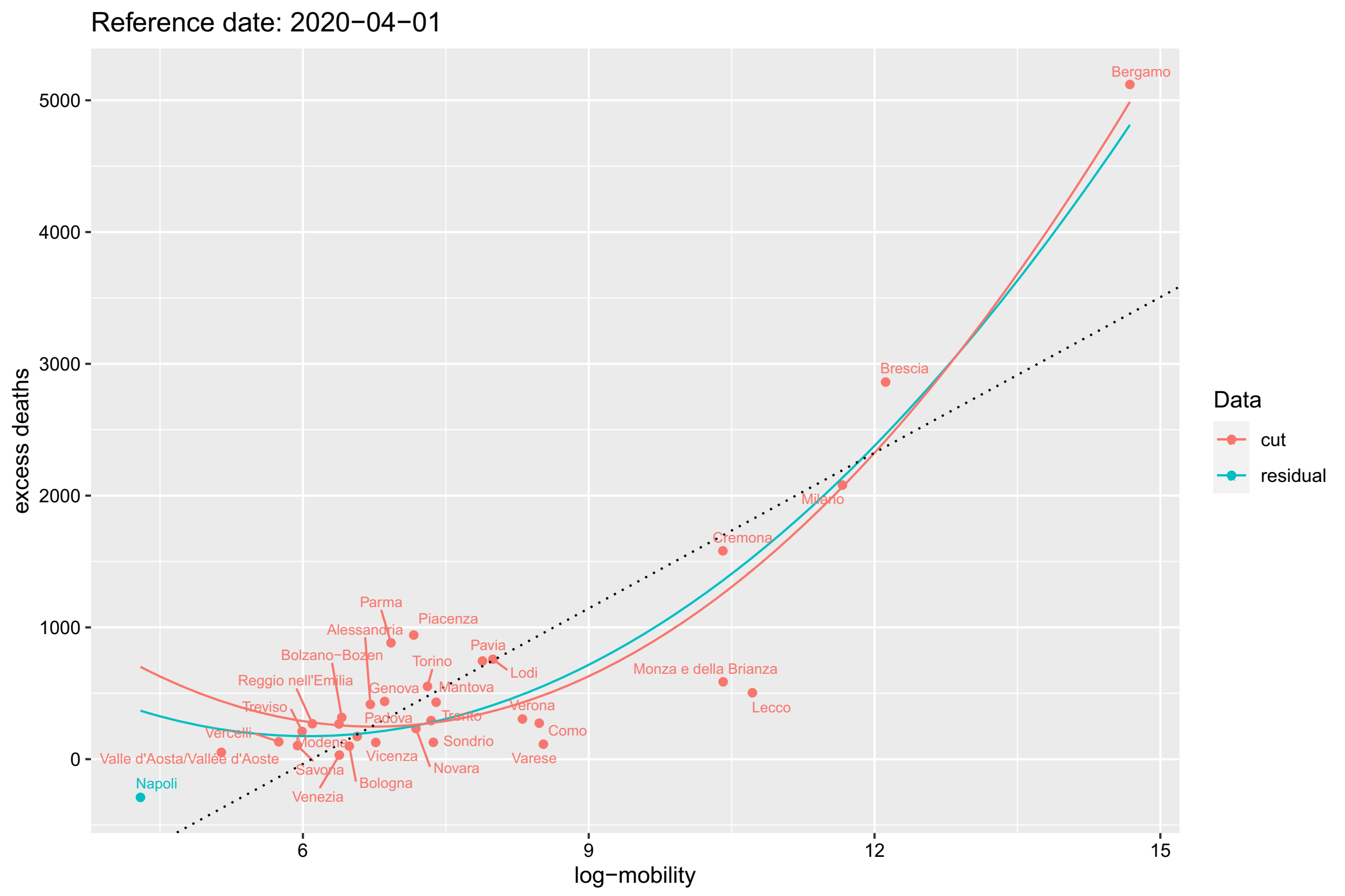

**Fig. 10** Data for log-mobility on 2 March 2020 and cumulative excess deaths on 1 April 2020. Red represents the selected subset of positive excess deaths and mobility. The other points are the residual data. Red and light blue together represent the whole data. The two curves correspond to the fitted model (2) for both data sets. The $R^2$ for this fitted model is 0.91. The straight line represent simple linear regression of the two variables in the plot. The black vertical lines are on the dates of local lockdowns on 21 and 27 February 2020 and country-wise lockdown on 9 March 2020. The green lines are the same as the black lines, but shifted by 14 days. (Color figure online)

for the times series $X_t$ and $Y_{t+\theta}$ and then to maximise it as a function of $\theta$. The lead–lag estimator $\hat{\theta}_n$ of $\theta$ is defined as [11]

$$\hat{\theta}_n = \arg \max_{-\delta<\theta<+\delta} |U_n(\theta)|.$$

When the value of $\hat{\theta}_n$ is positive, it means that $X_t$ and $Y_{t+\hat{\theta}_n}$ (or $X_{t-\hat{\theta}_n}$ and $Y_t$) are strongly correlated, so we say '$X$ leads $Y$ by an amount of time $\hat{\theta}_n$', so $X$ is the *leader* and $Y$ is the *lagger* and vice versa for negative $\hat{\theta}_n$. The lead–lag estimator is provided by the `yuima` R package [12]. Applying the lead–lag analysis to our data, we estimate a lag of 14 days which is in line with the median lag time from the onset of symptoms to death [18,24]. After shifting the normalised mobility curve by 14 days (`lagnmob`), the cross-correlation between the curve `lagnmob` and the curve `cumdeaths` raises from 0.69 to 0.98. A simple linear regression among these two series gives a value of $R^2$ of 0.96 (from 0.48 before shifting).

## 3.2 Correlation network analysis

After analysing the internal mobility of the Haut-Rhin department, we consider the internal mobility of each department and estimate a lead–lag parameter against the `cumdeaths` curves of all departments. This way of proceedings takes into account possible spin-offs from one department to another, but it is not a causal inference analysis. After running the extensive 9216 ($96 \times 96$) lead–lag analyses, we summarise the results through a network analysis. Therefore, we build a directed graph representing the lead–lag relationships

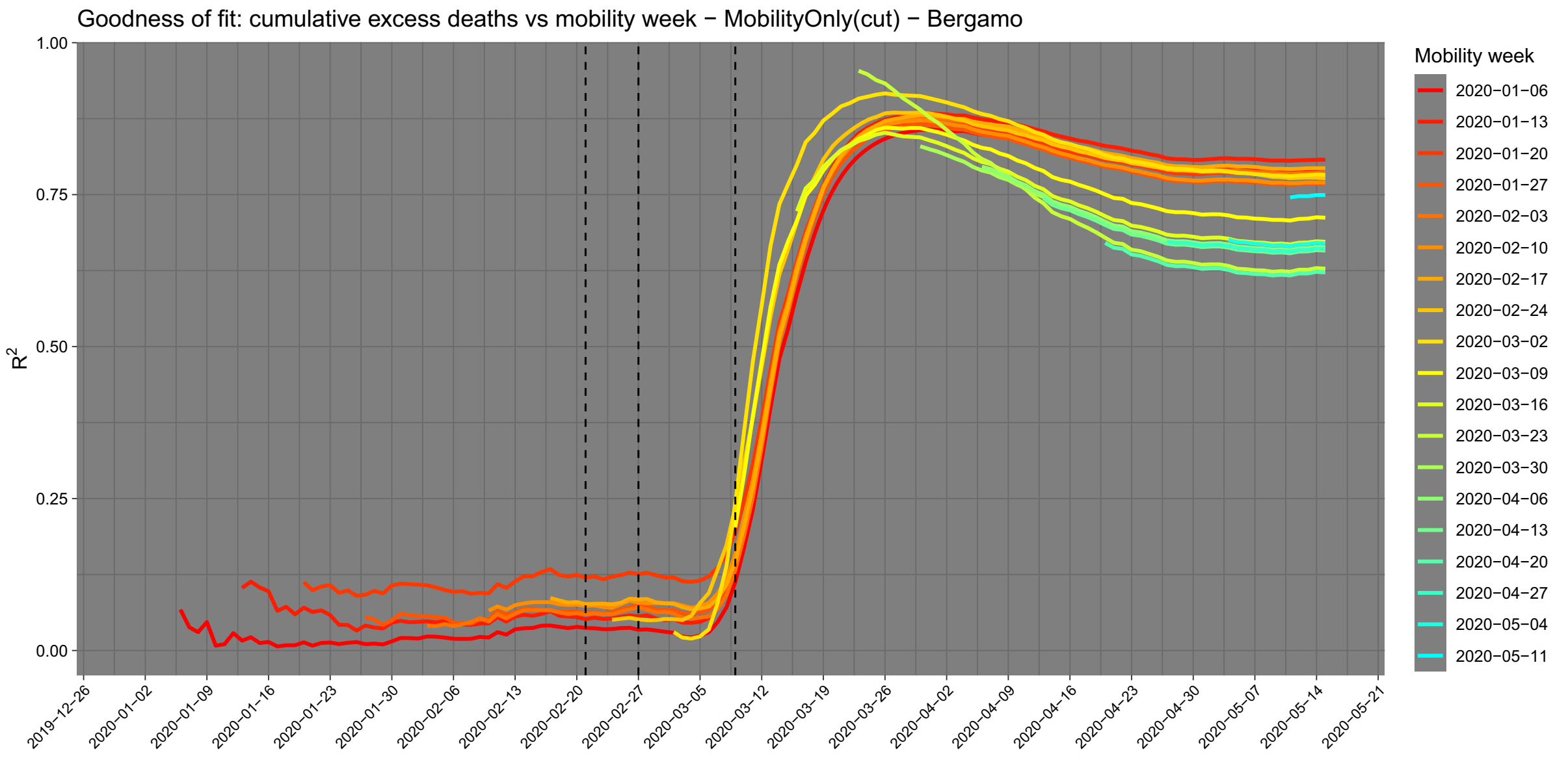

**Fig. 11** $R^2$ evolution for each mobility week using the model based on mobility alone on the selected data set of positive excess deaths and mobility index. The peak of correlation is on 23 March 2020 similarly to Fig. 5 with $R^2 = 0.95$

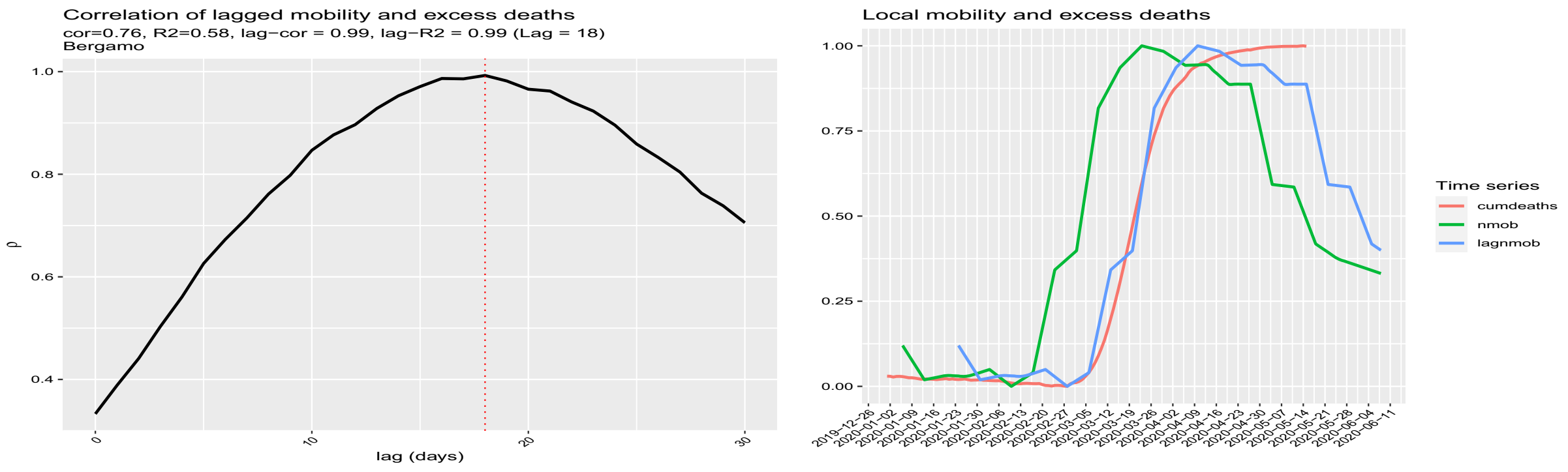

**Fig. 12** Left panel: Correlation between `cumdeaths` and `nmob` at different lags. Right panel: `cumdeaths`, `nmob` with the additional lagged normalised mobility curve `lagnmob`. The estimated optimal lag is 18 days

between the *lagger* cumulative deaths curves and the *leader* internal mobility. The nodes in the graph from $A_i$ to $B_j$ are such that $B_j$ is the `cumdeaths` curve for a target department $j$ and $A_i$ is the internal mobility of a department $i$. Cutting those edges that form loops (internal mobility on the same department) and further selecting the edges $i \to j$ which presents the maximal lagged correlation for a given destination $j$, we obtain a simplified graph. The edges on the graph are weighted according to the $R^2$ statistics. On this graph, a standard community detection algorithm for directed graphs is run in order to discover similar paths between mobility within a department and impact on other departments. Keeping in mind that lead–lag analysis does not involve any causal effect, the clusters that are obtained can be seen only as similar patterns of lagged cross-correlation between internal mobility and excess deaths in outbound departments. Most of the lag discovered are around 14–20 days. Shorter lag estimates usually correspond to very flat excess cumulative deaths curves. Too large lags (i.e. 30 days) are mostly spurious correlation due to few observations. Figure 7 shows a graphical representation of the network graph. (The full network is shown in Fig. 8.) It is important to stress that no causality effect should be interpreted from this correlation network graph. In fact, notice that the reason why the department of Hautes-Alpes is the origin of the edges of the graph towards many other departments is because the mobility there was reduced much earlier than in other departments, and therefore, the correlation with the cumulative excess deaths is much anticipated by simple covariation effects. So what can be retained from this analysis? This network says that dynamics of both mobility reduction and cumulative excess deaths have similar patterns regardless of the region. Therefore, this result, coupled with previous analysis on internal mobility, can inform policy makers on the expected spread containment in terms of human mobility reduction and the corresponding lag needed to achieve the flattening of the excess deaths curve.

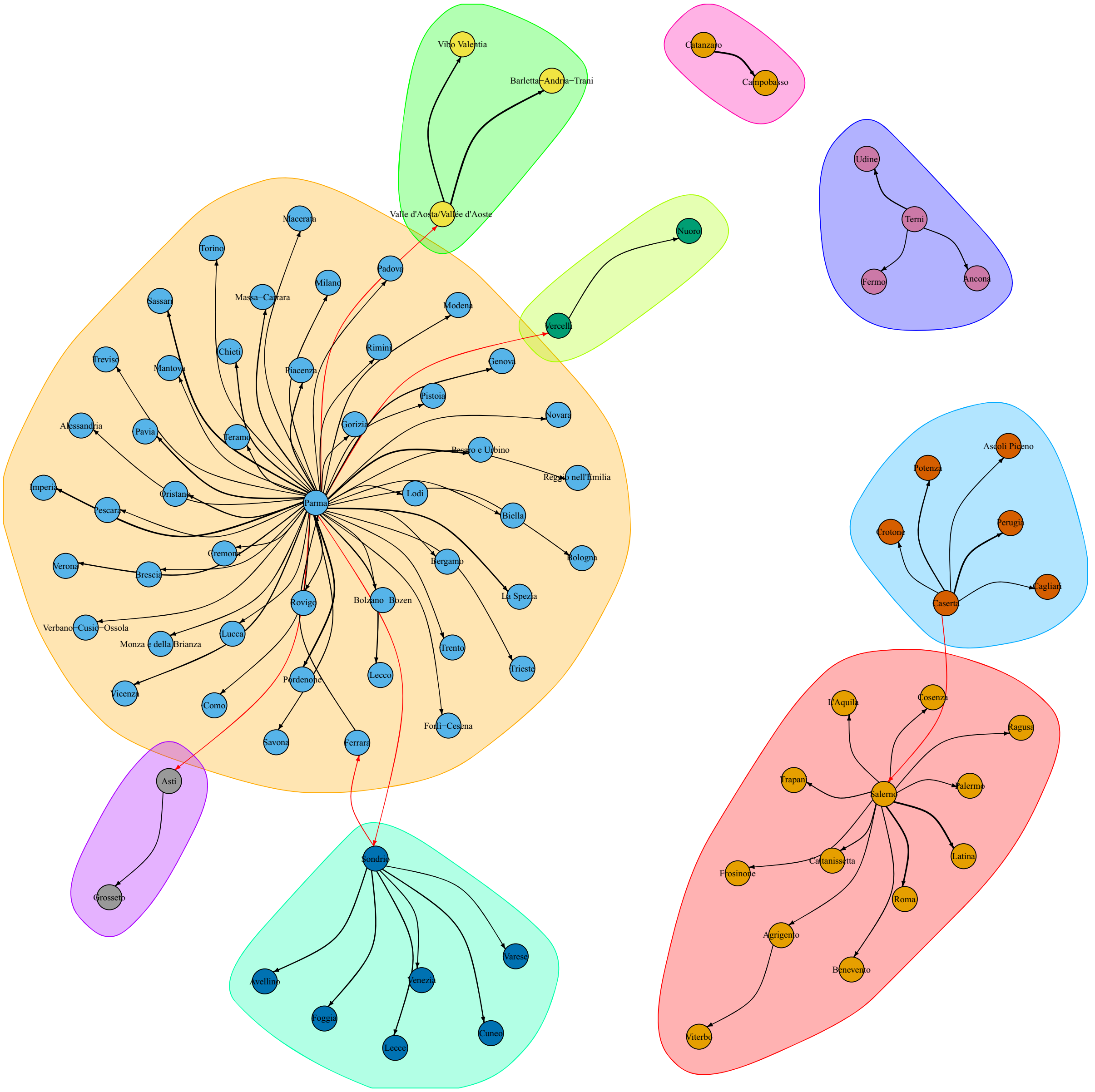

**Fig. 13** Full graph of lead–lag relationships and cluster of patterns. The width of the edges is inversely proportional to the estimated lag. Red edges connect two different clusters. No causality effect should be read from this graph. (Color figure online)

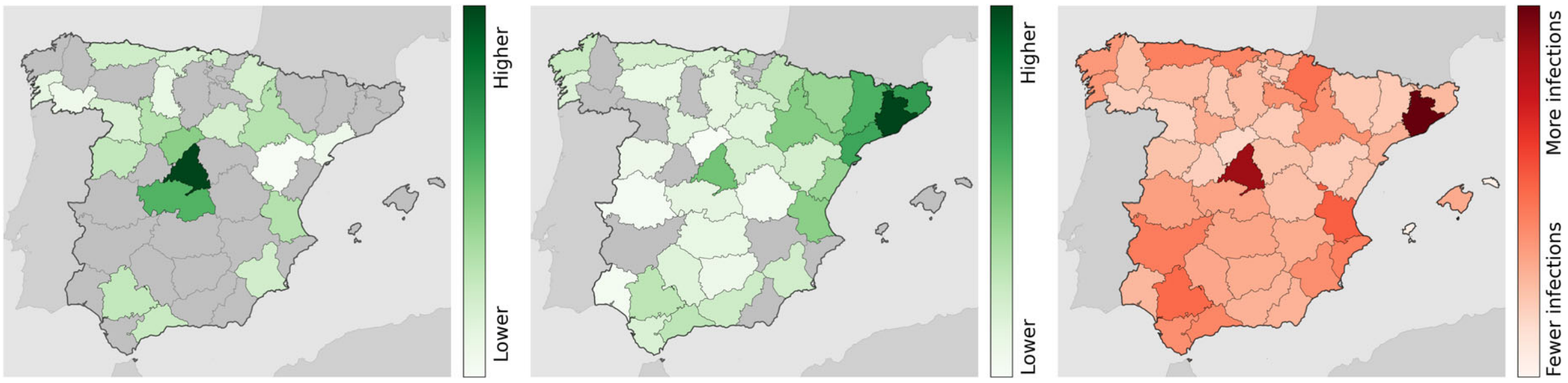

**Fig. 14** Left and middle panel: Mobility from Madrid and Barcelona for the week of 2 March 2020. Right panel: Total number of IgG positive cases

## 4 Human mobility and COIVD-19 spread in Italy

In this section, we have replicated the study of Sect. 3 for Italy. Data on excess deaths are publicly available[3] at Istat [15] at municipality level for 7270 municipalities (over a total of 7904, about 92%) for the period 1 January 2020–15 May 2020 (see Fig. 9). Mobility data for Italy are coming from two different MNOs at different resolutions in space (census areas and NUTS3) and time (hourly and daily). The use of both data sets allows to capture short movements (less than the hour, mainly internal mobility or across neighbour provinces) and longer ones (more than 1 h) through connectivity matrices even when aggregated at NUTS3 level as considered in our analysis. One of the provinces strongly hit by the virus initial spread in Italy is Bergamo. We will use this case to show a parallel analysis to that of Sect. 3. Figure 10 shows the scatterplot for the mobility week of 2 March 2020 against the excess deaths on 1 April 2020. Under the mode model (3), the $R^2 = 0.91$ for these two sets of data. Italy has put in place lockdowns for the red zones (very small municipalities) on 21 and 27 February 2020, and the country-wise lockdown was enforced on 9 March 2020, so it is interesting to see the pattern of correlation 14–20 days after 9 March 2020 as marked with vertical lines in Fig. 11.

Figure 11 shows that this evidence does not change substantially if we choose a different mobility week. Indeed, the figure shows that the $R^2$ evolution for each mobility week using the model (3) on the selected data set (`cut`) of positive excess deaths and mobility index present a common pattern. In this particular case, the $R^2$ reaches its maximum on 23 March 2020 with the value of 0.95. Now focusing on local mobility via lead–lag analysis, we discover a 18 days lag between mobility reduction and the cumulative curve of excess deaths (see Fig. 12). After shifting the normalised mobility curve by 18 days (`lagnmob`), the cross-correlation between the curve `lagnmob` and the curve `cumdeaths` raises from 0.76 to 0.99. A simple linear regression among these two series gives a value of $R^2$ of 0.99 (from 0.58 before shifting). The lead–lag analysis shows that, over the 110 Italian provinces, the lagged effect has a mean and median of 20 days, with the first quantile being 16 and third quantile 25 days. Figure 13 presents the correlation network graph for the Italian provinces. In Sect. 3, correlation is not causation.

## 5 Human mobility and IgG antibody testing in Spain

Another interesting case to test the impact of human mobility on the virus spread is Spain. Indeed, despite the fact that the number of cases is known only at region (NUTS2) or country level, this country has conducted a large-scale IgG SARS-Cov-2 antibody screening on the population at province level in two waves on 27 April 2020 and 11 May 2020. The study, conducted by ENE [4], has recruited 60,983 inland and an additional 3234 participants in the islands. As running examples, we consider the provinces of Madrid and Barcelona, the two provinces with highest number of cases in Spain, and Badajoz (in Extremadura region) which is an average case. In our study, we combine the results of the 2 weeks of testing and we compare these data with the mobility data using the same models of the previous sections. Figure 14 shows the mobility from Madrid and Barcelona towards other provinces of Spain in com-parison with the number of positive IgG positive cases by province. The combination of the left and middle maps suggests the existence of a relationship between mobility and positive cases that we will now try to quantify through correlation analysis. Figure 15 shows the log–log plot of mobility versus IgG positive cases. The pattern of the relationship between the two variables is similar to what we observed for the excess deaths of France (Fig. 4) and Italy (Fig. 10).

[3] Data released by Istat on 18 June 2020.

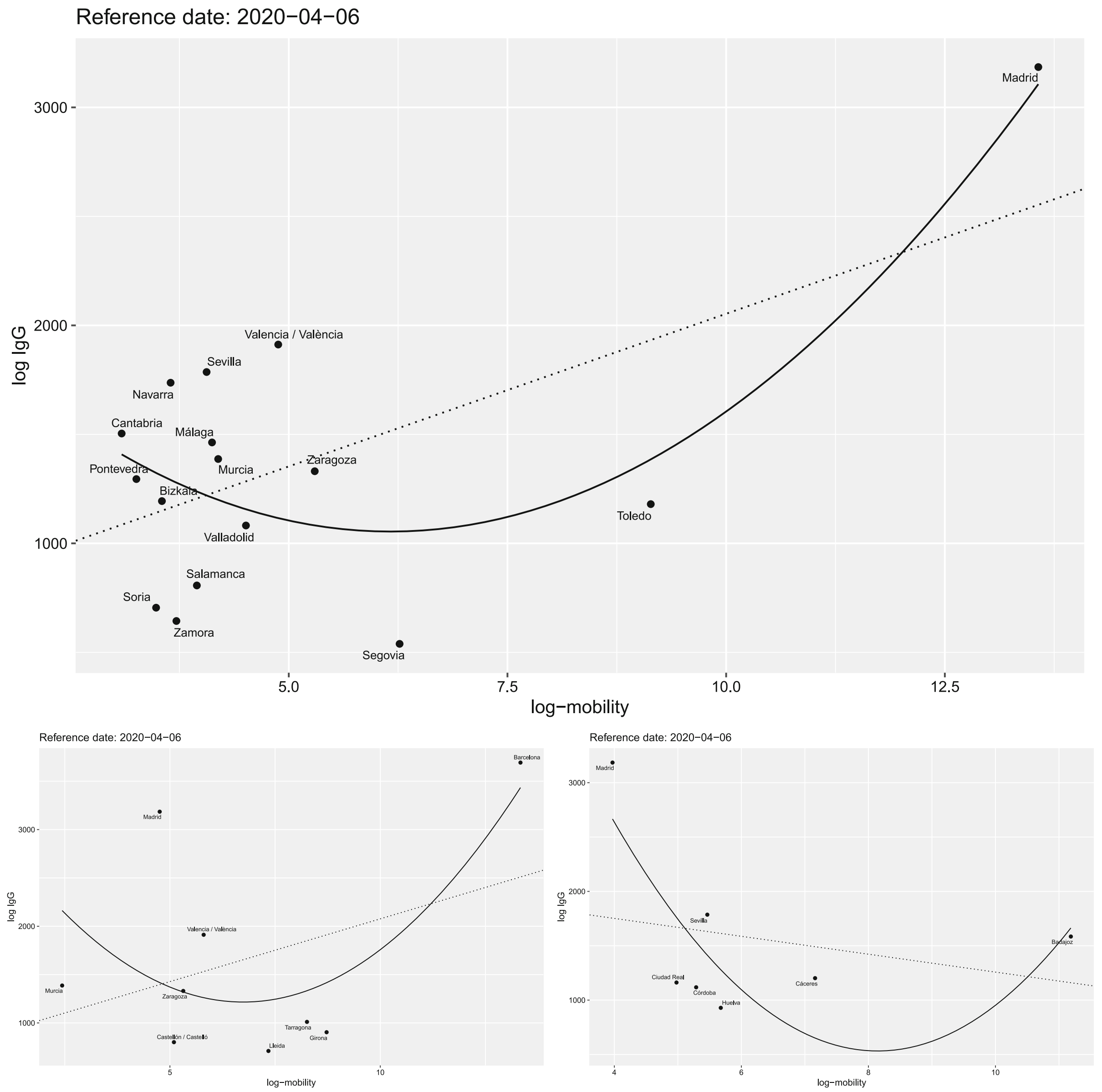

**Fig. 15** Data for log-mobility and total number of log IgG positive tests for the provinces of Madrid (top), Barcelona (bottom left) and Badajoz (bottom right). The shape of the relationship is similar to the cases of France (Fig. 4) and Italy (Fig. 10)

For Spain, though we do not have time series for the IgG tests, we cannot run a lead–lad or correlation network analysis like in Sects. 3 and 4, but we can only perform a simple correlation analysis, letting the time of mobility to vary and keeping the IgG data fixed. Figure 16 shows the behaviour of the Spearman (traditional) and Pearson (rank) correlation coefficients between the IgG positive tests and both the mobility and distance between provinces on the linear scale. The evi-dence here is that mobility is more important, in terms of simple correlation, than the proximity between the provinces up to lockdown which was enforced on Saturday 14 March 2020. The maximal value of the correlation coefficients for Madrid, Barcelona and Badajoz are, respectively, 0.67, 0.56 and 0.56. On the other side, Fig. 17 shows the goodness of fit measure $R^2$ for the (2) model on the log–log scale. For Madrid and Barcelona, $R^2 = 0.57$, while for Badajoz $R^2 = 0.62$.

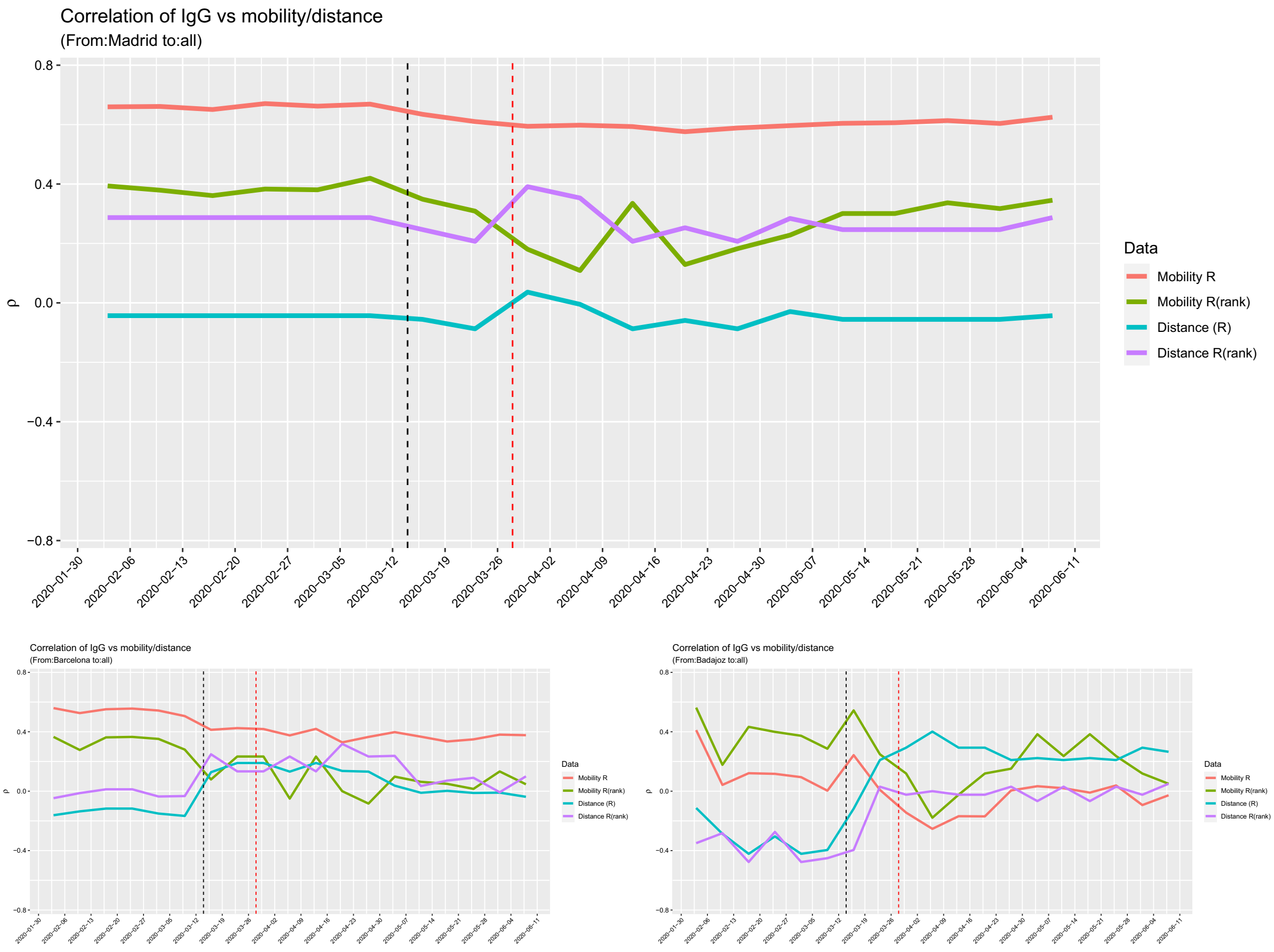

**Fig. 16** Values of the correlation coefficients for the provinces of Madrid (top, maximum $|\rho(\text{Mob})| = 0.67$), Barcelona (bottom left, maximum $|\rho(\text{Mob})| = 0.56$) and Badajoz (bottom right, maximum $|\rho(\text{Mob})| = 0.41$), between IgG positive tests and both mobility and distance. The black vertical line corresponds to the lockdown of 14 March 2020; the read line is just 14 days later

Clearly, the correlations are less strong than those of the excess deaths data because IgG data measure a different aspect of the COVID-19 spread and, begin essentially one point in space, is hard to estimate any variation. Still, the results confirm that human mobility have an impact on the virus spread.

## 6 Caveats of the study

It is important to remind that the choice to use a simple model of Eq. (2) aims at showing the rough effect of mobility on the initial spread of the virus. Clearly, such a model is not intended to and should not be used to produce any other type of result, like, for instance, to forecast the number of deaths. In this work, the excess deaths in France and Italy are treated only as a proxy for the virus spread, as well as IgG testing for Spain is a proxy for the number of people that have been in contact with the virus at a given time.
It is also well known that the excess deaths actually due to coronavirus are influenced by a long series of factors such as the age structure of the population and its health

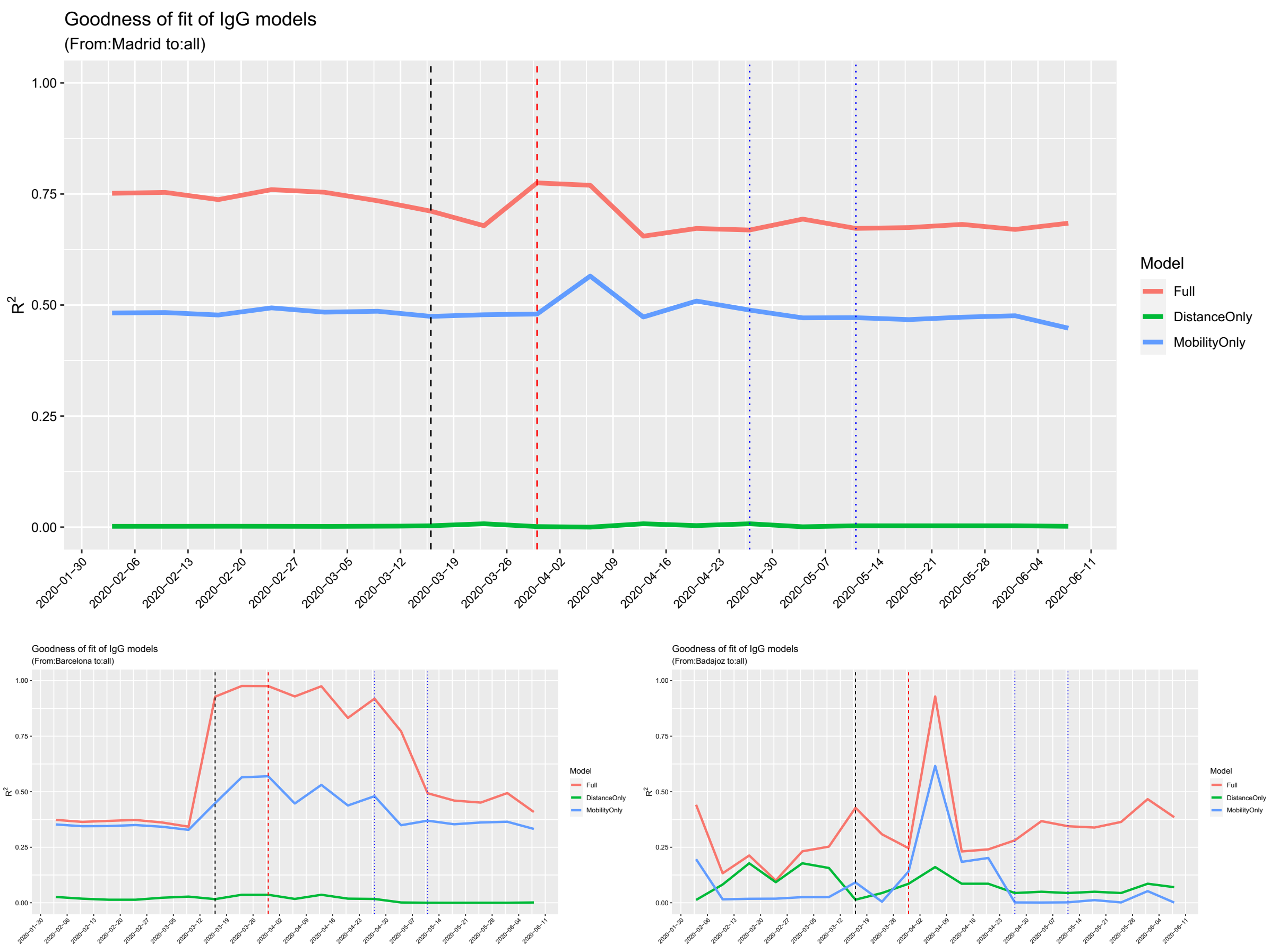

**Fig. 17** $R^2$ values for the provinces of Madrid (top, maximum $R^2 = 0.57$), Barcelona (bottom left, maximum $R^2 = 0.57$) and Badajoz (bottom right, maximum $R^2 = 0.62$), for the fitted model (2) between IgG positive tests and both mobility and distance. The black vertical line corresponds to the lockdown of 14 March 2020, and the red line is just 14 days later. The blue lines represent the interval of dates of the IgG screening. (Color figure online)

condition, the availability of ICU units and the preparedness of the healthcare system, and that these generally vary across regions. Moreover, part of the excess deaths after lockdown events, may also be deflated by the lower number of deaths due to, for example, road accidents or work-related (many activities have been shut down during national lockdowns). There is also a technical issue due to the construction of the ODMs where all movements below a given threshold are discarded in order to maintain data anonymity.

For obvious reasons (*in primis* for the lack of reliable information), all these aspects have not been accounted for by this study. In other words, despite some of the confounding effects are captured by the constant coefficient of the basic models adopted, clearly there are many other confounders that are not.

Yet, the fact that the results are quite stable and consistent across countries (also adopting different spatial granularities), together with very similar findings published from different research teams relatively to China [16], are absolutely encouraging for the evolution of this research.

## 7 Discussion

This study demonstrates that human mobility, derived by mobile data, is highly correlated with the spread of COVID-19 in the initial phase of the outbreak. In the case study of France, we have found that mobility can explain from 52 up to 92% of the excess deaths reasonably linked to the COVID-19 outbreak. In the case study of Italy, we have found similar results ($R^2$ up to 0.91 and lagged effect of 14–20 days, depending on the provinces), confirming that human mobility has a high impact on the virus spread, at least before other physical distancing measures are in place. In the case of Spain, we have found that the number of people resulting positive to IgG tests is highly correlated ($\rho$ up to 75%) with the human mobility.

The above case studies can provide solid evidence to forecasting scenarios for future waves of the virus, in cases where only limited additional protective measures are in place (e.g. wearing masks, physical distancing, etc.). This is achieved by exploring the linkage between the geographical distribution of the excess of mortality with respect to 2019 and fully anonymised and aggregated mobile positioning data from European MNOs. Besides predicting dynamics of future COVID-19 outbreaks, connectivity information can be used as basis to plan targeted control measures to curb the spread of the virus. Future data gathered in the context of this European Commission initiative with MNO could enable an analysis at EU regional scale, providing a framework for sharing best practices and data-driven input to inform coherent strategy among EU Member States for de-escalation and recovery.

**Acknowledgements** The authors acknowledge the support of European MNOs (among which A1 Telekom Austria Group, Altice Portugal, Deutsche Telekom, Orange, Proximus, TIM Telecom Italia, Telefonica, Telenor, Telia Company and Vodafone) in providing access to aggregate and anonymised data, an invaluable contribution to the initiative. The authors would also like to acknowledge the GSMA[4] (GSMA is the GSM Association of mobile network operators.), colleagues from DG CONNECT[5] (DG Connect: The Directorate-General for Communications Networks, Content and Technology is the European Commission department responsible to develop a digital single market to generate smart, sustainable and inclusive growth in Europe) for their support and colleagues from Eurostat[6] (Eurostat is the Statistical Office of the European Union) and ECDC[7] (ECDC: European Centre for Disease Prevention and Control. An agency of the European Union) for their input in drafting the data request. Finally, the authors would also like to acknowledge the support from JRC colleagues, and in particular the E3 Unit, for setting up a secure environment, a dedicated *Secure Platform for Epidemiological Analysis and Research* (SPEAR) enabling the transfer, host and process of the data provided by the MNOs; as well as the E6 Unit (the Dynamic Data Hub team) for their valuable support in setting up the data lake.

**Data and materials availability** Any request of MNO data and derived products should be agreed with each operator, owner of the data. Data on excess deaths are publicly available from INSEE [14]. Data on excess deaths are publicly available from Istat [15]. Data on the IgG SARS-Cov-2 antibody screening in Spain available from ENE [4]. The analysis was entirely conducted with base R with the additional package `yuima`. R code is available upon request to the authors.

**Compliance with ethical standards**

**Conflict of interest** The authors declare that they have no conflict of interest .